# Surface Self-Assembly of Functionalized Molecules on Ag(111): More Than Just Chemical Intuition

Andreas Jeindl[1], Jari Domke[2], Lukas Hörmann[1], Falko Sojka[2], Roman Forker[2], Torsten Fritz[2], and Oliver T. Hofmann[1]*

[1] Institute of Solid State Physics, NAWI Graz, Graz University of Technology, Petersgasse 16, 8010 Graz, Austria

[2] Institute for Solid State Physics, Friedrich Schiller University Jena, Helmholtzweg 5, 07743 Jena, Germany

**ABSTRACT:** The fabrication of nanomaterials involves self-ordering processes of functional molecules on inorganic surfaces. To obtain specific molecular arrangements, a common strategy is to equip molecules with functional groups. However, focusing on the functional groups alone does not provide a comprehensive picture. Especially at interfaces, processes that govern self-ordering are complex and involve various physical and chemical effects, often leading to structures that defy chemical intuition, as we showcase here on the example of a homologous series of quinones on Ag(111). From chemical intuition one could expect that such quinones, which all bear the same functionalization, form similar motifs. In salient contrast, our joint theoretical and experimental study shows that profoundly different structures are formed. Using a machine-learning-based structure search algorithm, we find that this is due to a shift of the balance of three antagonizing driving forces: adsorbate-substrate interactions governing adsorption sites, adsorbate-adsorbate interactions favoring close packing, and steric hindrance inhibiting certain otherwise energetically beneficial molecular arrangements. The theoretical structures show excellent agreement with our experimental characterizations of the organic/inorganic interfaces, both for the unit cell sizes and the orientations of the molecules within. With a detailed examination of all driving forces, we are further able to devise a design principle for self-assembly of functionalized molecules. The non-intuitive interplay of similarly strong interaction mechanisms will continue to be a challenging aspect for the design of functional interfaces. Our agreement between theory and experiment combined with the new physical insights indicates that these methods have now reached the necessary accuracy to do so.

Many properties of thin films, such as optical properties[1] or electrical conductivity,[2] are determined by the structure they assume upon adsorption on the substrate.[3] To engineer functional interfaces, it is therefore imperative to understand and predict which structures form for a given material combination.[4–8] At a single molecule level, relevant handles to influence their properties are well known. A typical example are conjugated organic molecules, which are relevant for organic nanoelectronics.[9–11] There, increasing the π-electron backbone or introducing functional groups systematically affects optical properties.[12–15] At the same time, changing either the backbone or the functional groups will also change the molecule's crystal polymorphs and its physical properties[16–19] in non-obvious ways. Particularly for thin films, and even more so for monolayers, the complex interplay between intermolecular and molecule-surface interactions can lead to the formation of new packing motifs.[3]

A starting point to design complex adsorbate layers with specific properties is to exploit diverse chemical design principles based on chemical intuition. Previously probed design principles involve non-covalent interactions,[20–29] halogen bonding,[30–33] dipole-dipole interactions[34–38] or steric blocking[39] and shape complementarity.[23,28,29,32] When one of these interactions is dominant, an intuitive guess of the resulting motifs can be made. Unfortunately, at interfaces the interplay with the surface often thwarts this approach. Therefore, a typical approach to design such molecules is to combine chemical intuition with increasing empirical knowledge from preceding experiments, enabled by molecular-resolution scanning tunneling microscopy (STM). The driving forces leading to self-assembly can then, in hindsight, be analyzed by experimental and theoretical methods. However, typically, even when the driving forces for a specific system are known, a holistic design of particular motifs is still prevented by the fact that the knowledge of the driving forces cannot be easily transferred from one case to another. Quite contrarily, even systems that have similar interactions can form disparate structures, as we demonstrate hereafter.

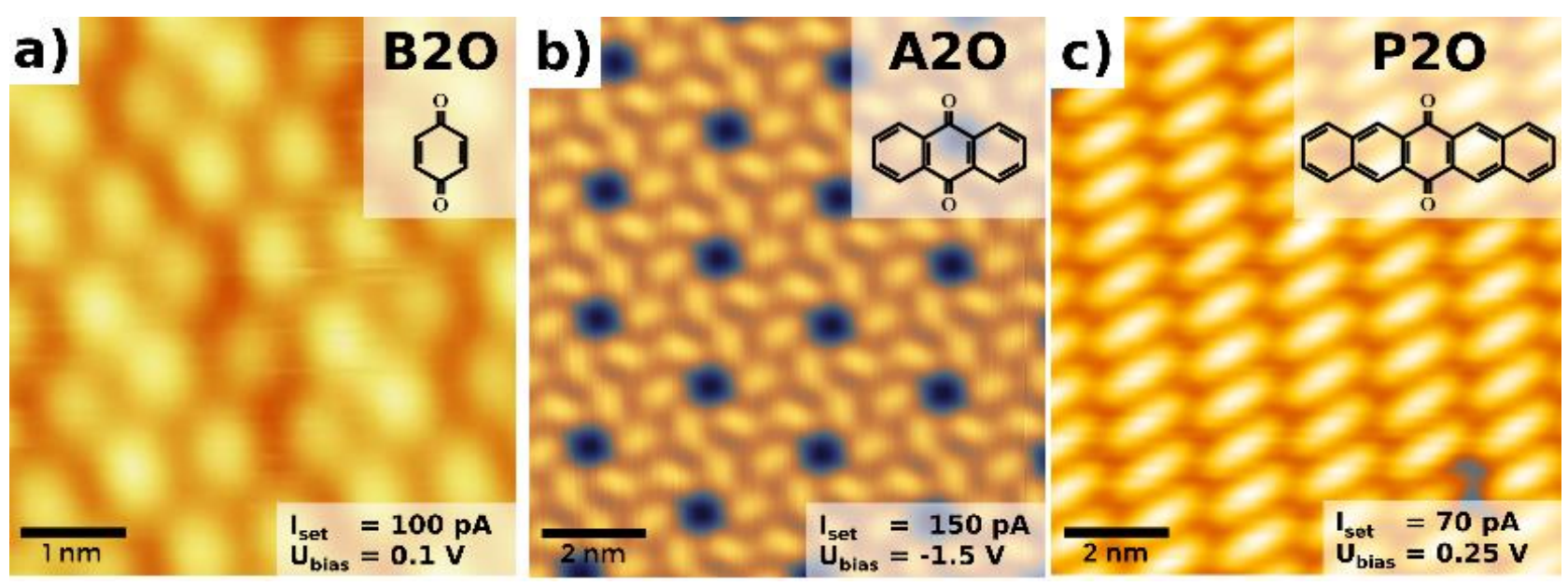

**Figure 1. Comparison of experimental constant-current scanning tunneling microscopy (STM) images for molecular monolayers of a) benzoquinone (B2O); b) anthraquinone (A2O); and c) pentacenequinone (P2O) on a Ag(111) surface prepared by physical vapor deposition in ultra-high vacuum.**

In this work, we examine the predictability and transferability of driving forces for on-surface molecular structures. For this we use a homologous series of molecules with identical functionalization, namely the quinones 1,4-benzoquinone (B2O), 9,10-anthraquinone (A2O), and 6,13-pentacenequinone (P2O). In all three molecules, the functional groups provide directed forces for self-assembly via (i) highly attractive intermolecular interactions between the oxygens and hydrogens, and (ii) a strong molecule-substrate interaction due to surface-induced aromatic stabilization.[40] Conversely, the backbone interacts non-site-specific via van der Waals forces. When depositing up to a single monolayer on Ag(111), we find three entirely different structures (overview in Figure 1; a detailed characterization is given later in this work). While the smallest molecule, B2O, exhibits a simple 2D surface pattern, the larger A2O forms symmetric hexagonal rings with voids in between. The largest molecule of this series, P2O, crystallizes in close-packed molecular rows, as also found by others.[40,41] Despite the chemical similarity, the backbone size thus decisively determines which motifs form.

We provide systematic insight into how the backbone size affects the driving forces leading to the formation of these motifs. To this end we predict, based on first principles, which on-surface motifs the three quinones form on Ag(111), utilizing a combination of density functional theory (DFT) and machine learning (details see Methodology). First, we focus on the interactions of individual molecules with the Ag(111) surface to unveil trends in the molecule-substrate interaction. We then investigate the intermolecular interactions of close-packed molecular layers on the Ag surface. Mapping the intermolecular energies onto specific molecule parts enables us to identify the main contributors for polymorph formation. This procedure allows extracting general trends for those interactions. We also find that the aspect ratio of the molecules plays a large (and hitherto probably undervalued) role, determining how many favorable interactions with neighboring adsorbates a single molecule can obtain. Finally, a detailed examination of all driving forces for the experimentally observed structures allows to devise a design principle for the applicability of functional groups to tailor molecular self-assembly on surfaces.

## RESULTS AND DISCUSSION

### Interaction of Individual Molecules with Ag(111)

Generally, individual quinone-functionalized molecules interact with the Ag(111) surface by site-specific interactions between the oxygens and the metal, as well as non-site-specific van der Waals interactions between backbone and substrate. One could thus expect that in the absence of intermolecular interactions all three different quinones prefer the same or similar adsorption geometries on the metal (defined by the adsorption site on the metal substrate and azimuthal rotation of the molecule). To test whether this is indeed the case, we first performed a pre-screening of a potential energy surface (PES) with reduced dimensionality, followed by full geometry optimizations (details in Methodology).

To compare the three molecules, we focus on three important aspects (Figure 2): (i) the adsorption geometries, (ii) the adsorption energies, and (iii) their energy distribution (i.e., how many adsorption geometries exist within a certain energy range). Despite different backbone lengths, the adsorption geometries are rather similar, while the number of different adsorption geometries increases with backbone size. As a consequence, all adsorption geometries of B2O and A2O feature a corresponding adsorption geometry of P2O, where the oxygens are approximately in the same positions. Figure 2a illustrates this point, showing the P2O geometries sorted by their adsorption energy and overlaying the corresponding B2O and A2O geometries.

Although the adsorption geometries are very similar for the three molecules, their energetic ranking is very different (Figure 2b). This suggests a key-lock-like interaction between the quinones and the surface. Presumably, which geometries are stable is governed (mostly) by the oxygens that bind to specific sites on the Ag(111) surface. Conversely, their energetic ranking depends on the registry of the backbone with the surface. For the most stable geometries, the adsorption energy increases with increasing backbone size (-1.11 eV for B2O, -1.50 eV for A2O, -2.15 eV for P2O), as expected.

As introduced above, the second factor determining which motif forms is the intermolecular interaction. An energetically unfavorable adsorption geometry might still be part of the energetically most favorable motif if it accommodates more attractive intermolecular interactions compensating the loss in adsorption energy.

For our systems, the range of adsorption energies decreases with increasing molecule size from 300 meV for B2O to 170 meV for P2O. Concomitantly, an increasing number of stable adsorption geometries is found energetically close to the most stable geometry, making more adsorption geometries easily accessible for monolayer formation. Especially for P2O, already 8 of its 12 adsorption geometries are found within a range of 50 meV. In general, making the molecule larger by adding benzene rings to the quinone backbone increases the number of adsorption geometries and increases adsorption energies. Simultaneously, the energy difference between different minima decreases, leading to a weaker adsorption-geometry dependence. This is a direct consequence of the inherent incommensurability of the acene backbone with the Ag(111) surface: the larger the backbone, the more the energetic landscape of the molecule-substrate potential becomes smoothed out.

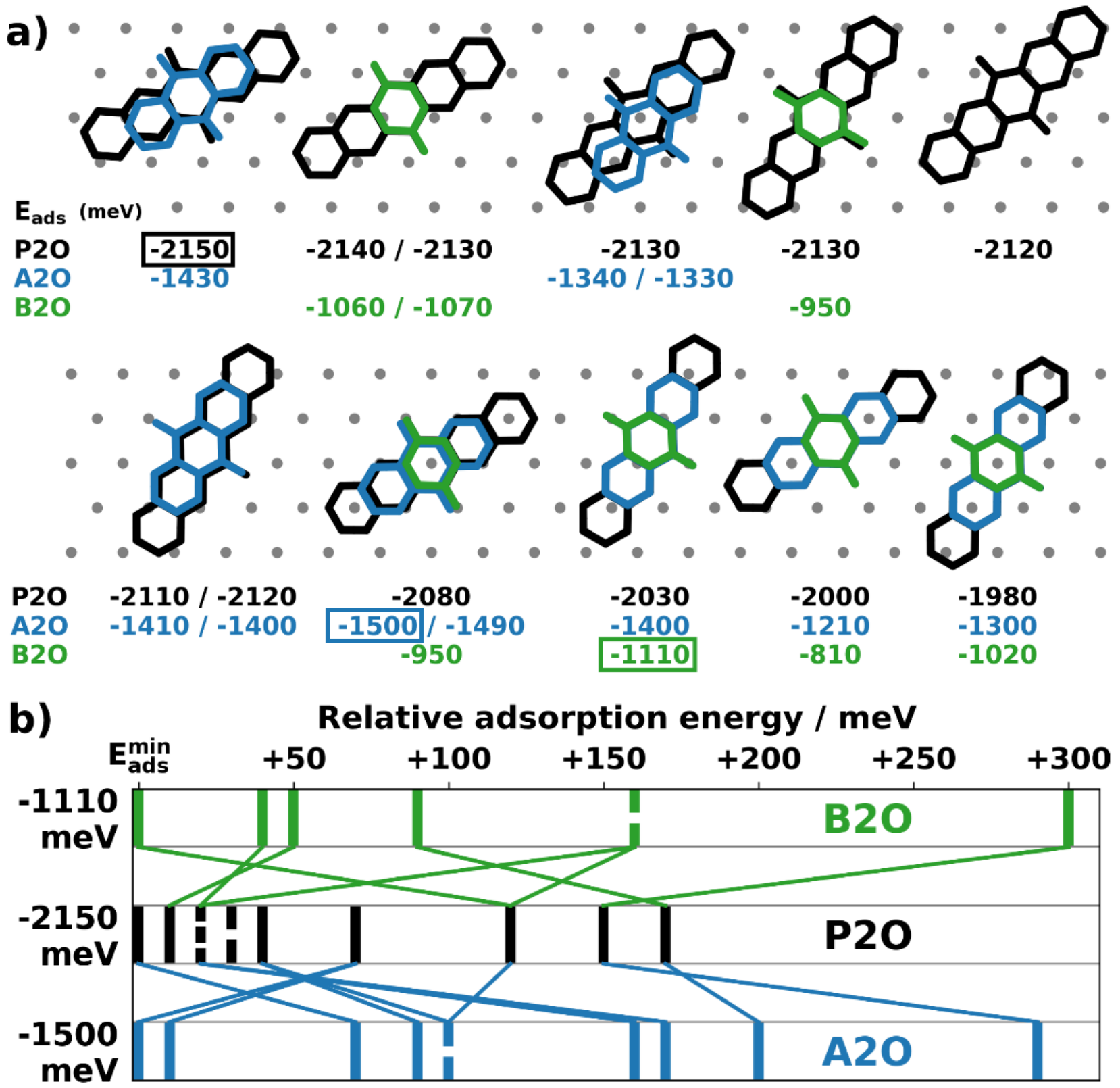

**Figure 2. a) Visualization of all symmetry-inequivalent adsorption geometries for B2O, A2O and P2O molecules with their corresponding adsorption energies (negative values of $E_{ads}$ denote energy gain upon adsorption). Surface substrate atom positions are indicated with grey dots. Boxes mark the energetically best adsorption geometries. All energy values are given in meV. Two energies for a single visualized geometry mark hcp- and fcc-hollow sites, respectively (i.e., there are two adsorption geometries which only differ due to stacking of the second substrate layer). There, the left energy corresponds to the geometry shown here (hcp hollow site). b) Spread of all adsorption geometry energies for all three molecules relative to the energy of the respective best geometry. Adsorption energies with multiple energetically equivalent geometries are indicated with dashed lines. There, the number of dashes is equal to the number of geometries. The colored connecting lines indicate the energetic reordering compared to the P2O geometries.**

## Intermolecular Interactions on the Surface

Now that we understand interactions of individual molecules with the surface, we proceed to the intermolecular interactions on the surface. The number of possible arrangements of molecules on the surface, however, is enormous.[42] This makes an exhaustive mapping of the interactions intractable. A similar challenge is encountered for first-principle structure determination, where interactions are approximated using intermolecular interactions from gas phase data[43] or via modified force fields.[44] Those methods perform best for low coverages or weak molecule-surface interaction. Here, however, the focus is on close-packed structures with strong molecule-surface interactions, rendering those methods inapplicable.

For this reason, we used the SAMPLE[45] approach. It uses the previously mentioned adsorption geometries as constituents to build a large but discrete set of potential motifs with (here) up to six molecules per unit cell and various unit cell sizes (details in Methodology). From the millions of potential motifs, we selected the approx. 250 most diverse candidates using D-optimality[46] and calculated their formation energies using DFT. These calculations were then used to infer all relevant molecule-substrate and intermolecular interactions on the surface using the energy model (Equation 1) with Bayesian linear regression.

$$E = \sum_{geoms} N_g U_g + \sum_{pairs} N_p V_p \quad (1)$$

Here, $U_g$ is the adsorption energy of a molecule with the adsorption geometry $g$, and $V_p$ is the interaction between every pair of molecules (called "pairwise interaction" hereafter) in the motif. $N_g$ and $N_p$ denote how often the corresponding interactions appear in each motif. With this method we can predict the formation energies for all potential motifs with a leave-one-out cross validation error of less than 20 meV per adsorbate molecule. Figure 3a gives an overview over the intermolecular interaction energies ($V_p$) for the three investigated systems. The most attractive interactions lead to an energy gain of up to 200 meV for a pair of B2O molecules, 250 meV for A2O and 300 meV for P2O. Hence, for larger molecules also the intermolecular interactions become stronger. We also estimated the energetic contributions of the interactions between the adsorption-induced dipoles for all tight-packed motifs (details in Methodology) The interaction energies of those dipole sheets are all below 6 meV per molecule, rendering adsorption-induced dipole-dipole interactions a negligible factor for the systems presented.

Besides the total interaction energies within the motifs, the specific form of the energy model also allows to extract and visualize all pairwise interactions used in the model. Figure 3b shows these interactions spatially resolved for B2O, A2O and P2O. There, the molecule in the center is kept fixed and the second molecule is moved (at fixed rotation) to all different possible adsorption sites around it. Each circle indicates the center of the second molecule with the color of the circle corresponding to the interaction energy of this molecular pair. The discretization of the pairwise interactions stems from the usage of adsorption geometries as fixed building blocks.

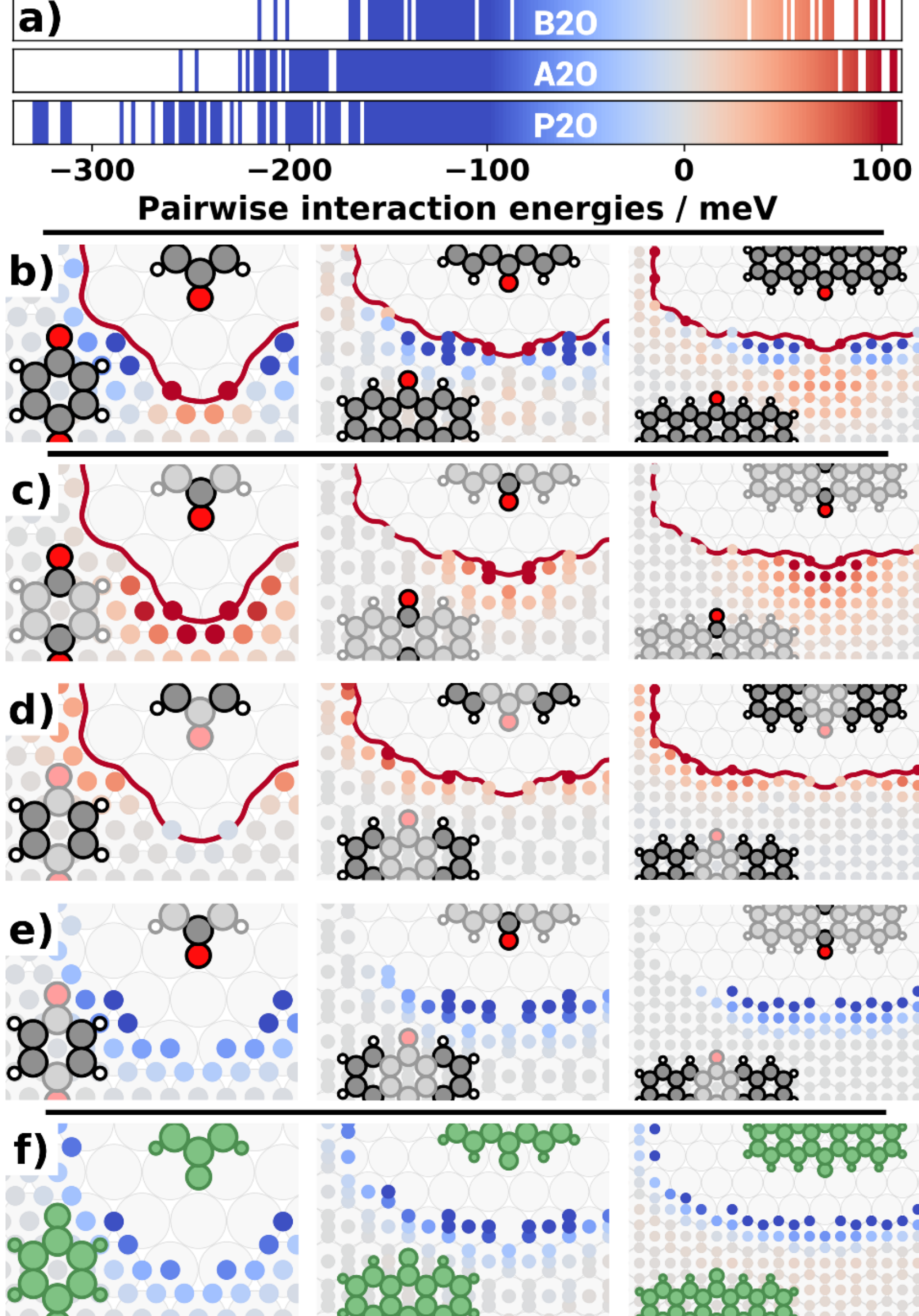

**Figure 3. Visualization of pairwise intermolecular energies and their contributions. a) Distribution of interaction energies for the three molecules under investigation. Colored areas represent interactions present in the given energy window; b) Total pairwise interaction energies for the three quinones. Each circle represents a possible interaction between the central molecule and another molecule centered at the circle position. The red contour shows the minimal distance before a pair is considered colliding. The circle color (same color scale as in a) indicates the corresponding interaction energy; c)-e) electronic interaction energies between molecules mapped onto different molecule parts: c) oxygen - oxygen d) hydrogen - hydrogen e) oxygen - hydrogen interactions; f) van der Waals interactions between molecules**

The general form of interactions is similar for all three systems and independent of the size of the backbone. For a deeper insight, we investigate the influence of the molecular fragments by breaking up $V_p$ into a sum of fragments $V_p^f$ (Equation 2).

$$E = \sum_{geoms} N_g U_g + \sum_{pairs} \sum_{fragments} N_p^f V_p^f \qquad (2)$$

This allows us to map intermolecular interactions onto specific parts. Figure 3c-e show the interactions between oxygens (c), between carbon rings (d), and the interactions between oxygens and rings (e). To separate electronic effects and van der Waals contributions, we performed this mapping on fragments without the van der Waals contributions to the energy. The van der Waals contributions are shown in a separate panel (Figure 3f).

The intermolecular oxygen interactions (Figure 3c) are exclusively repulsive, following Coulomb-like behavior due to the partially negatively charged oxygen atoms on both molecules. Interactions between rings (Figure 3d) are dominated by the proximity of hydrogen atoms and are also purely repulsive. The only attractive electronic interactions occur between carbon rings and oxygens (Figure 3e). The second, highly attractive, contribution stems from van der Waals interactions between molecules (Figure 3f). This is insofar surprising, as the molecules are all flat lying on the surface, which results in a relatively small contact area between adjacent molecules. While those results are qualitatively consistent with chemical intuition, we resolve the interactions based on chemical groups in a quantitative and position-specific way, revealing the general intermolecular interaction characteristics for a homologous series of quinones.

### Stable Motifs in Theory and Experiment

Having characterized all relevant interactions on the surface, we can now evaluate which motifs are expected to be observed. For this, we evaluated the energy model (equation (2)) for all the millions of candidates and identified the best motifs in terms of formation energy per molecule. The details of the prediction process are given in Supporting Section 3. In the following, we focus on the structures predicted to be energetically most favorable and their comparison to the experimental thin films prepared via physical vapor deposition in ultra-high vacuum (details in Methodology). We note that the phases shown in Figure 1 (and discussed here in more detail) are not the only structures that form upon deposition. A detailed description of the full polymorphism is beyond the scope of this manuscript and will be provided elsewhere.

The qualitative agreement between theory and experiment can already be seen in Figure 4a-d. To obtain a quantitative insight, we compare both, the experimental low energy electron diffraction (LEED) images and diffraction patterns obtained via fast Fourier transformation (FFT) of the scanning tunneling microscopy (STM) images (details in Supporting Section 2), with kinematic scattering simulations for the predicted structures, taking into account the tabulated structure factors for all atoms (see Methodology). Figure 4e-h compares the experimental diffraction patterns and deduced unit cells for B2O, A2O and P2O to our best-fitting low-energy predictions.

P2O exhibits multiple different motif candidates within the prediction uncertainty. All of them contain the same molecular rows observed in experiment. However, our model allows for various relative arrangements of these rows (see Supporting Figure S7). The best structure with parallel rows contains a single molecule per unit cell and an area of 125.9 $Å^2$/molecule. It is in excellent agreement with the experimentally deduced structure, not only within the STM image (Fig 4d) but also with respect to spot positions and intensities (Figure 4h). The lattice lengths agree perfectly for the long and within 4 % for the short axis. The enclosed angle agrees within 5° (Table 1).

The experimental preparation of A2O led to well-ordered structures exhibiting a periodic hexagonal pattern. As can be seen in the STM image (Figure 4c) the experimental surface structure is in good agreement with the energetically most favorable theoretical structure containing six molecules per unit cell with an area of 88.6 Å$^2$/molecule. This cell shows excellent agreement with the predicted structure within fit uncertainties (Figure 4g).

For B2O, we predict the energetically best motif (labeled Motif I) to contain two non-equivalent molecules per primitive unit cell. The molecules in this cell are oriented in parallel but located at different adsorption sites, i.e., they are (slightly) nonequivalent. Within SAMPLE, we also find energetically low-lying defects where rows of molecules are rotated by 90°, which, within our model, costs 100 meV per defect while allowing for a denser packing of the B2O molecules. For the sake of discussion, we also consider a limiting case where every other row consists of these defects. This structure (Figure 4j), which will be called Motif II hereafter, contains four non-equivalent molecules. For comparability we also use an equivalent cell with four molecules for Motif I (Figure 4i). The interpretation of the experimental diffraction pattern (shown in blue in Figure 4f, details explained in Supporting Section 2.2) indicates that B2O exhibits a line-on-line[47] registry. Converting that periodicity into real space (Figure 4j) shows that the computed and experimental lattice vectors differ only by -6 % and +8 %, respectively, while the unit cell areas differ only by 2 %. The enclosed angle is reproduced within 6°. We attribute these minor variations to the fact that the calculations require periodic boundary conditions which artificially and unavoidably enforce commensurability between the adsorbate and the substrate, which is different from the line-on-line registry in the real motif.

A summary of the numerical values for the discussed cells is presented in Table 1. More details for all cells shown in Figure 4 are given in Table S2. The agreement between our first-principles structure search and the experimentally found motifs underlines the fidelity of our analysis.

Table 1: Comparison of experimental (grey background) and theoretical unit cells.

| | $a_1$ [Å] | $a_2$ [Å] | Γ [°] | θ [°] | A [Å$^2$] |
|---|---|---|---|---|---|
| B2O* | 13.8 | 11.4 | 88.5 | -14.3 | 157.4 |
| | 13.0 | 12.4 | 94.3 | -10.9 | 160.9 |
| A2O | 24.6 | 24.9 | 120.0 | -36.5 | 530.0 |
| | 24.8 | 24.8 | 120.0 | -36.6 | 531.6 |
| P2O | 15.0 | 8.2 | 95.9 | 40.8 | 121.3 |
| | 15.0 | 8.5 | 100.9 | 40.9 | 125.9 |

a1, a2: Lengths of lattice vectors derived from the epitaxy matrices in combination with the theoretical substrate lattice vectors (primitive lattice constant of Ag(111) 2.842 Å); Γ: Angle between lattice vectors; θ: Angle between first lattice vector and primitive substrate axis; A: Unit cell area.*Data for Motif II, for other unit cells see Table S2

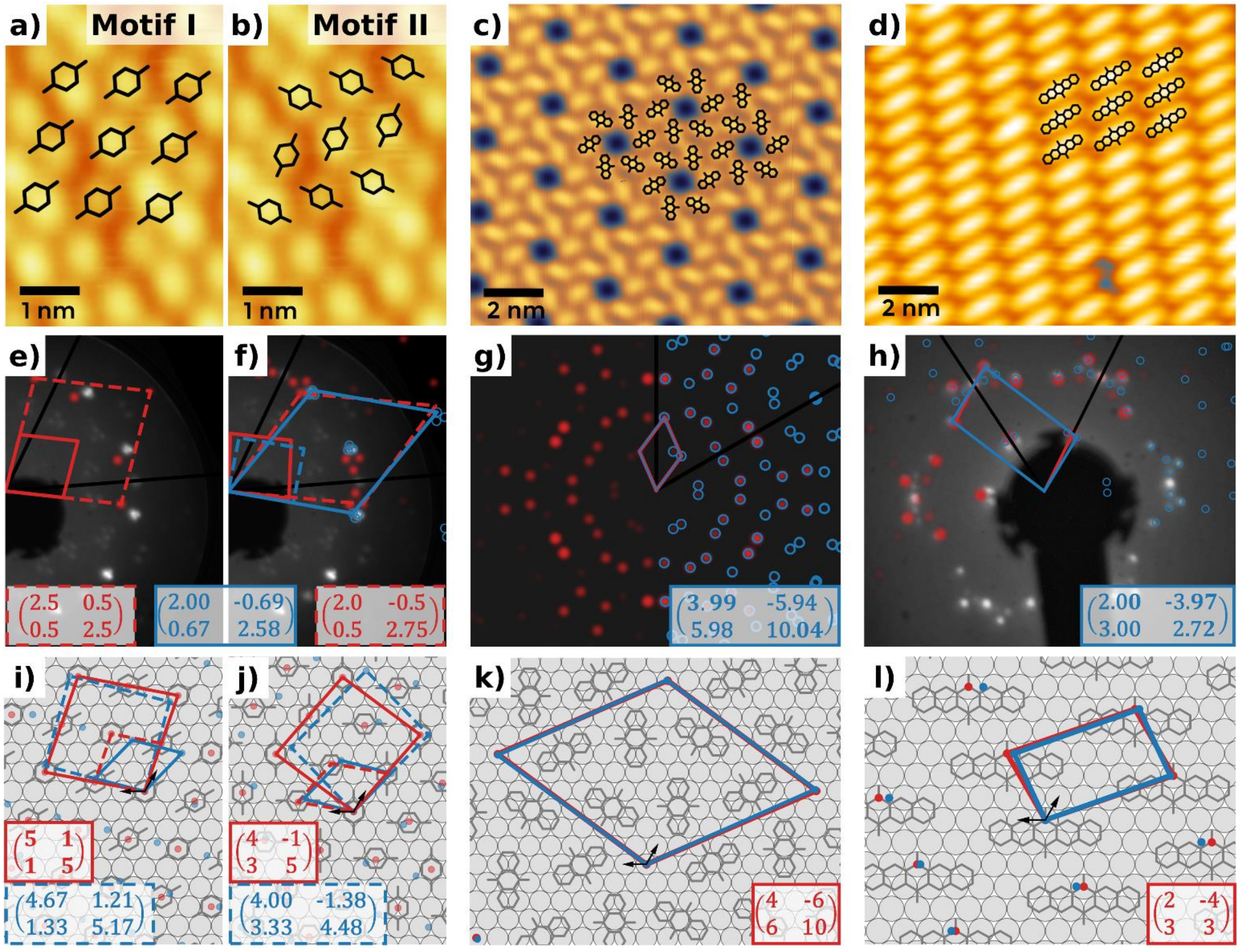

**Figure 4. Comparison of theoretical findings (red) with experimental results and interpretations thereof (blue). a-d) Comparison of STM experiments (see Figure 1) with theoretically predicted surface polymorphs. e-h) Comparison of theoretical LEED patterns obtained via kinematic diffraction theory (red) to fits of FFTs from STM images (blue) and LEED images; primary electron energies are 27 eV for B2O and 48 eV for P2O. i-l) Visualization of the real-space on-surface arrangement. The epitaxy matrices represent the unit cells in the respective substrate basis given by the black arrows. Fit uncertainties for the experimental epitaxy matrices (blue) are below 0.08 for all elements (details in Supporting Table S2). For B2O, two possible theoretical phases are shown. While LEED experiments reveal the primitive adsorbate unit cell, simulations for B2O require 4 molecules in a commensurate supercell. The dashed unit cells in e-f) and i-j) therefore represent transformed cells to obtain comparability (experimental cells replicated; simulated cells reduced).**

### Influence of Steric Hindrance

We have shown earlier in this work that the pairwise interactions as well as the molecule-surface interactions are similar for all three different systems. Nevertheless, the motifs observed experimentally and predicted theoretically exhibit substantially different features. This shows that there must be a crucial, hitherto missing, factor influencing the packing motifs formed. This factor is the steric hindrance between molecules. To illustrate and quantify this effect, we took the energetically most favorable pair for each system and evaluated the interactions to a third molecule at different orientations and positions. The resulting visualization (Figure 5) is similar to Figure 3b, but now the orientation of the outer molecule is visualized via the orientation of a rectangle. All rectangles were scaled according to their absolute energy to focus on stronger interactions. All rectangles were colored according to their energy. Furthermore, the size was scaled according to their absolute energy to focus on stronger interactions. Note that here the energy range is ± 200 meV to focus on the strongest trimer interactions. The visualization reveals that B2O can arrange such that the energetically most favorable interactions in all directions can be exploited for monolayer formation, leading to highly attractive interactions in four directions for each molecule. A2O can build structures with molecular triangles, which also allow highly attractive interactions per molecule, as can be seen on the lower right corner of Figure 5b. For P2O, the conjugated backbone is so long that the attractive hydrogen-oxygen interactions can only be formed with two neighbors. This makes interactions involving molecules rotated relative to each other energetically much less favorable, leading instead to long rows of molecules aligned parallel to each other.

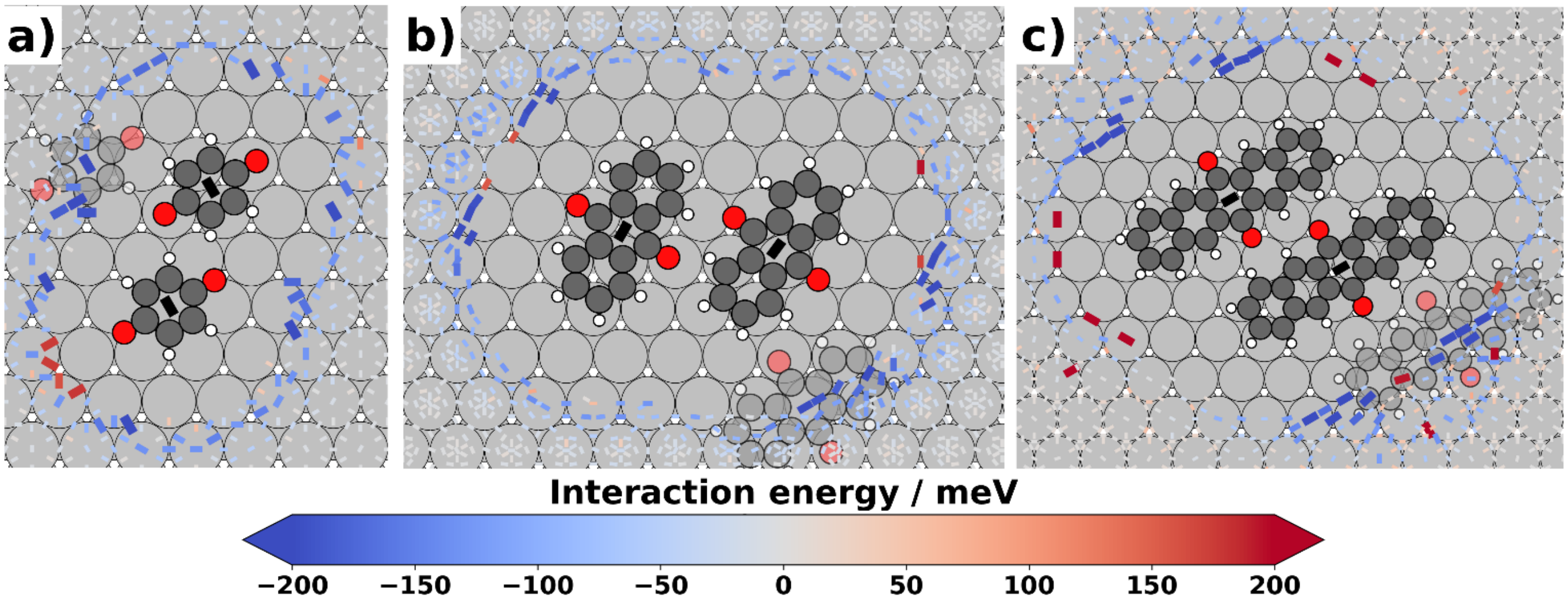

**Figure 5. Trimer interactions visualizing all possible interactions of the energetically best pair with an outer molecule for a) B2O, b) A2O and c) P2O. Each rectangle represents a molecule centered at the rectangle's midpoint, with its backbone orientation indicated by the long axis of said rectangle. The respective rotation indicates the energetically most favorable third molecule at this specific point, all other rotations at each site were discarded for clarity. The rectangles are scaled according to their absolute energy value to highlight strong interactions. The third, semi-transparent molecule indicates an energetically favorable geometry.**

### Evaluating the Driving Forces

With all driving forces at hand, it is possible to quantitatively discuss the energetic contributions for the motifs formed by each molecule (Figure 4 i-l). Figure 6a shows the mean molecule-substrate interactions of the observed motifs compared to the corresponding best possible molecule-substrate interaction for a single molecule on the surface. In Figure 6b we compare the intermolecular interaction energies for the observed motifs (green arrows) to the hypothetically best possible intermolecular interactions obtainable when fully neglecting molecule-surface interactions (shaded areas).

As a first step, the energy decomposition now allows us to briefly reconsider the energetic differences between the two B2O motifs we compared to experiment. For both motifs, the molecule-substrate interactions are roughly the same when averaged over all comprising adsorption geometries. The energetic difference is the result of intermolecular interactions. Thus, within the commensurate model, Motif I is the favored structure in terms of *effective* energies (green arrows in Figure 6b). The decomposition, however, shows that van der Waals and oxygen-hydrogen interactions (*cf.* blue arrows) are slightly more favorable for Motif II. Yet, its effective intermolecular energy is less favorable than that of Motif I due to the stronger oxygen-oxygen repulsion marked by red arrows. The energetic difference between Motifs I and II might decrease or even vanish if Motif II were to exhibit a slightly incommensurate structure, which would allow it to decrease O-O repulsion by slightly increasing the distance between rows with different molecule rotation. While ab-initio methods are currently not able to predict incommensurate structures, our energy decomposition here allows to anticipate the implications of slight incommensurability.

We can now continue with a comparison of the driving forces for the three different molecules. For A2O the observed motif (Figure 4) includes only the energetically best adsorption geometries, while the corresponding motifs for B2O and P2O also contain worse-ranked local geometries, resulting in a small adsorption energy penalty. This penalty is compensated by the larger intermolecular interaction energies those two systems can realize compared to A2O. While for all four motifs the role of van der Waals interactions is similar, the electronic contributions of the fragments decrease with increasing molecule size. This is caused by a reduced availability of highly attractive intermolecular interactions due to steric hindrance. The interaction energy for the best motif of A2O is lower than for P2O due to the tradeoff between molecule-substrate and intermolecular energies it needs to take. This could be reformulated into a design principle:

*For molecules to form structures dominated by functional-group-induced intermolecular interactions, the molecule-substrate energy corrugation (represented by the number of low-energy adsorption geometries) should be less than the energy to be gained by intermolecular interactions.*

This design principle is fulfilled for the largest molecule in the series, P2O, due to low substrate corrugation and also for the smallest molecule, B2O, due to the highly favorable intermolecular interactions. However, it is not fulfilled for the intermediate-sized A2O, where the molecule-surface interaction is more prominent than the intermolecular driving forces. One could a priori assume that, if such a design principle is fulfilled for one molecule with a given functionalization, it would also be fulfilled for others with the same functionalization, but here we see that this is not even the case for a homologous series. This highlights the importance of input from first-principles calculations when applying such design principles.

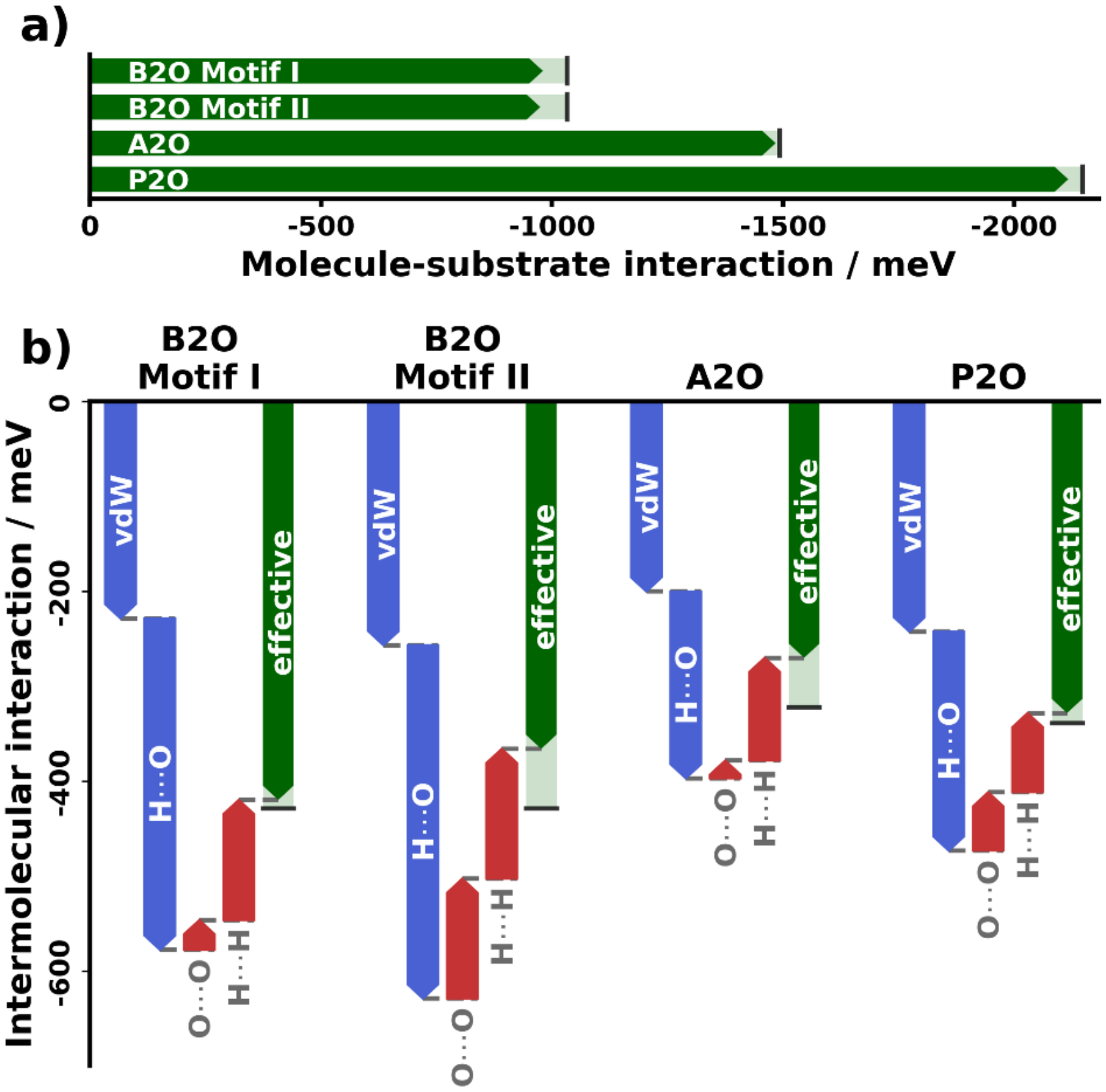

**Figure 6: Detailed breakdown of interaction energies for the motifs formed by the three molecules (Figure 4 i-l) into van der Waals (vdW) and electronic interactions between molecular fragments. a) Molecular adsorption energy. b) Total intermolecular interaction energies per molecule separated into contributions of vdW and molecular fragments. The green arrows indicate the resulting interaction energy. The green shaded areas indicate the hypothetically best possible values for the contributions if they could be realized separately, which is not possible as real motifs are always a tradeoff between molecule-substrate and intermolecular interactions.**

## CONCLUSION

In this work, we addressed the question why – in salient contrast to intuition – molecules with identical functionalization form completely different structural motifs on a metal surface, and whether insights and general trends for the on-surface self-assembly mechanisms can be retrieved from first principles without prior experimental input. To understand this behavior and reveal general trends, we investigated all the interaction mechanisms involved.

For individual molecules on the surface, the increasing molecule size increases the number of local minima while decreasing the energetic corrugation. Nonetheless, the adsorption geometries occurring for all molecules, are similar. The main intermolecular interactions driving the motif formation in quinones are a balance of van der Waals interactions and attractive interactions between oxygen and hydrogen atoms counteracted by repulsive oxygen-oxygen and hydrogen-hydrogen interactions. The relative strengths of these interactions differ between the three investigated molecules, but not sufficiently to explain the vastly different surface structures. The more varying factor is the availability of those interactions in tight-packed structures: The compact structure of B2O allows four oxygen-hydrogen interactions per molecule. These interactions lead to a brick-wall structure. For A2O, the larger backbone favors the formation of triangular building blocks, which, in combination with highly beneficial adsorption geometries, lead to the observed hexagonal ring-like motif. The even larger P2O molecules can only interact favorably with two neighboring molecules. The anisotropy of these interactions leads consequentially to the parallel alignment of P2O, thereby forming molecular rows. Our findings are backed by a remarkable agreement between theoretically predicted structures and experimentally resolved packing motifs.

These insights showcase that the vastly different surface patterns are driven by a changing balance of surface-molecule and intermolecular interactions in combination with steric hindrance upon increasing the molecule size. Even for relatively simple systems like a homologous series of quinones it is therefore challenging, if not outright impossible, to predict or engineer the monolayer structures based on chemical intuition alone. On a more positive note, advanced computational tools based on machine-learning, such as our SAMPLE[45] approach, allow to retrieve quantitative interaction energies and extract general trends for the interaction mechanisms. With enough structures investigated, we expect this to pave the way for a more holistic design of surface structures.

## METHODS

All calculations were performed with the FHI-aims package[48] using the exchange-correlation functional PBE.[49] Long-range dispersion was included via the $TS^{surf}$ correction.[50,51] The integration in k-space was performed with a Γ-centered grid with a well converged density of 36 points per primitive lattice direction and one k-point in z direction. As our calculations involved unit cells with different shapes, the k-points were scaled according to the length of the unit cell vectors. The periodic nature of our systems allowed us to use the repeated slab approach with a unit cell height of 80 Å (including >50 Å of vacuum), a dipole correction,[52] and eight layers of Ag with a mixed-quality numerical basis set (details in Supporting Section 1.1). With this approach, all adsorption energies were converged to a methodological uncertainty below 20 meV per adsorbate molecule.

Finding all local minima for the molecules on the surface would, in principle, require an exhaustive global structure search. This is infeasible even for the most advanced algorithms[53,54] due to the high configurational complexity.[42] Thus, we performed a two-step procedure that starts by first optimizing a single molecule on the surface and consecutively utilizing the BOSS approach[54] to find all local extrema in the three-dimensional (x, y, and rotation around molecular axis) PES. As a second step, geometry optimizations were performed from all extrema in the aforementioned PES where the whole molecule as well as the two topmost metal layers were allowed to relax. All final adsorption geometries, where at least one atom position differed by more than 0.1 Å (with symmetries taken into account), were considered as separate adsorption geometries (see also Supporting Section 1.2).

We define the adsorption energy as $E_{ads} = E_{sys} - E_{sub} - E_{mol}$ where $E_{sys}$ is the energy of the combined system, $E_{mol}$ the energy of a molecule in the gas phase, and $E_{sub}$ the energy of the pristine Ag slab with the two upper layers pre-relaxed. Negative values of $E_{ads}$ denote energy gain upon adsorption.

The SAMPLE[45] approach takes the surface atom positions in a given unit cells as a discrete grid and generates all combinations of building blocks at all possible positions within the unit cell and then removes colliding structures. As building blocks, it uses all adsorption geometries with all their symmetry equivalents on the respective metal surface. To not only be limited to a single unit cell, with SAMPLE we also generate an exhaustive set of unit cells for a given unit cell size (number of substrate atoms). For all three molecules in this study we varied the unit cell sizes and number of molecules per cell to ensure that experimentally feasible motifs are part of the prediction set (details in Supporting Section 1.3). The training set for the SAMPLE approach was chosen with experimental design employing the D-optimality criterion[46] on interactions in the motifs. For the description of the different configurations within SAMPLE, the species dependent feature vector, considering distances between hydrogens and oxygens in all combinations, was used. All hyper parameters were thoroughly converged to robust values (Supporting Section 1.4). To further reduce the computational costs, we first calculated the training points as free-standing monolayers (i.e., removed the metal substrate) and used the resulting fit parameters for the pairwise interaction energies as priors for the on-surface systems. With this methodology we could reduce the number of needed training calculations to 249 for B2O, 245 for A2O and 84 for P2O.

The fully trained energy models were then used to predict the energies and rank the full set of possible motifs. Details of this prediction process can be found in Supporting Section 3.

The interaction energies of adsorption-induced dipoles with their periodic replicas are estimated by summing over all dipole interaction energies via $W = \frac{\mu^2}{4\pi\epsilon_0}\sum\frac{1}{r_i^3}$. Here, $r_i$ is the distance between the central unit cell and each of its neighbors and $\mu$ is the effective point dipole of the unit cell. The sum over unit cells is performed until the energy is converged to below $10^{-4}$ eV.

To compare the theoretical structure to LEED results, simulations based on kinematic diffraction theory were performed. Therefore the location and intensity of the peaks were calculated as the square of the structure factors $n_{\vec{G}} = \sum_{atoms} f_{atom}(\vec{G}) \exp(-i\vec{G}\vec{r}_{atom})$. Here $\vec{G}$ are the reciprocal lattice vectors of the crystal and $\vec{r}_{atom}$ the positions of atoms in the unit cell, respectively. The atomic form factors were each approximated with $f_{atom}(G) = \sum_{i=1}^{n} a_i \exp\left(-b_i\left(\frac{G}{4\pi}\right)^2\right)$ where $a_i$ and $b_i$ were taken from the international tables for crystallography(2006).[55]

The organic molecules B2O (CAS: 106-51-4, nominal purity 99.5%), A2O (CAS: 84-65-1, nominal purity 97%) and P2O (CAS: 3029-32-1, nominal purity 99%) were obtained as powders from Sigma-Aldrich. A2O and P2O could be further purified by temperature gradient vacuum sublimation using a CreaPhys DSU-05. B2O was purified in a home-built sublimation device consisting of two separate glass tubes, dubbed reservoir and sublimation tube, which are connected to each other by an angle valve, with the sublimation tube additionally attached to the deposition chamber by a dosing valve. B2O was initially filled into the reservoir tube and evacuated with both valves open. The reservoir tube was subsequently heated by a stream of hot air until a sufficient amount of B2O deposited on the walls of the sublimation tube, after which both valves were closed.

The Ag(111) single crystals were obtained from MaTeck GmbH and cleaned by $Ar^+$ sputtering at 700 eV and incident angles of ±45° to the surface normal, followed by annealing at 800 K. Sputtering and annealing were cyclically repeated until the surface quality was satisfactory, as confirmed by low energy electron diffraction (LEED).

The monolayers were deposited at room temperature (296 K) in an ultra-high vacuum chamber with a base pressure better than $5 \cdot 10^{-10}$ mbar via physical vapor deposition. The deposition of B2O was carried out by positioning the sample in approximately 30 cm distance and in direct line of sight to the dosing valve and opening it until the chamber pressure reached 1·10-6 mbar, after which the valve was kept open for 10 minutes. For the deposition of A2O, the purified powder was filled into another glass tube and connected to the deposition chamber via a dosing valve. During layer deposition the sample was placed in approximately 30 cm distance and in direct line of sight to the dosing valve, and the dosing valve was opened for 15 minutes. P2O was deposited by thermal evaporation from a shutter-controlled effusion cell held at 450 K with a deposition time of 10 minutes.

After deposition, the P2O samples were additionally gently heated while being monitored by LEED until a well-ordered structure became visible. For the quantitative structural analysis, the samples were first characterized by distortion-corrected LEED[56] at room temperature including a numerical fitting of the assumed surface unit cell in reciprocal space to the measured LEED pattern (LEEDLab 2018 version 1.4). After LEED examinations, the samples were transferred into a low-temperature scanning tunnelling microscope (SPECS JT-LT-STM/AFM with Kolibri Sensors) and measured in constant-current mode at 4.5 K for A2O and 1.2 K for B2O and P2O. Afterwards, the obtained STM images were subjected to a two-dimensional Fourier transform and the epitaxy matrices were determined from those (Details in Supporting Section 2), utilizing the same software tools that were used for the LEED measurements, as already described elsewhere.[57–59]

## ASSOCIATED CONTENT

### Supporting Information

The Supporting Information contains details of the experimental and theoretical methods, the fit procedure for the A2O unit cell, and additional pairwise interaction plots

### Data Availability

The calculations used in this manuscript are available via the NOMAD database (www.nomad-repository.eu). For the review process, they can be found as data sets authored by "Jeindl Andreas" under the names "X2O adsorption geometries" for the local adsorption geometry, "X2O structure search gasphase training data" for the prior data in gas phase, and "X2O structure search training data" for the training of the SAMPLE model. DOIs to the sets will be provided prior to publication.

Experimental details are provided in the Supporting Information; further experimental data is available upon request.

## AUTHOR INFORMATION

### Corresponding Author

Oliver Hofmann - Institute of Solid State Physics, NAWI Graz, Graz University of Technology, Petersgasse 16, 8010 Graz, Austria; orcid.org/0000-0002-2120-3259; Phone: +43 316873-8964; Email: o.hofmann@tugraz.at

### Notes

The authors declare no competing financial interests.

## ACKNOWLEDGMENT

We acknowledge fruitful discussions with M. Scherbela,A.T. Egger and E. Zojer. Funding through the projects of the Austrian Science Fund (FWF): P28631-N30 and Y1175, by the Deutsche Forschungsgemeinschaft (DFG): FR 875/16-1 and FR 875/19-1, and by the Federal Ministry of Education and Research of Germany (BMBF – KMU-NetC, 03VNE1052C) within the project "InspirA" is gratefully acknowledged. Computational results have been achieved in part using the Vienna Scientific Cluster (VSC) and using resources of the Argonne Leadership Computing Facility, which is a DOE Office of Science User Facility supported under Contract DE-AC02-06CH11357.

# Supporting Information to “Surface Self-Assembly of Functionalized Molecules on Ag(111): More Than Just Chemical Intuition”

A. Jeindl[1], J. Domke[2], L. Hörmann[1], F. Sojka[2], R. Forker[2], T. Fritz[2], and O.T. Hofmann[1*]

[1] Institute of Solid State Physics, NAWI Graz, Graz University of Technology, Petersgasse 16, 8010 Graz, Austria

[2] Institute of Solid State Physics, Friedrich Schiller University Jena, Helmholtzweg 5, 07743 Jena, Germany

*Corresponding author: o.hofmann@tugraz.at

# 1 Supporting Methodology (Theory)

## 1.1 Mixed Basis Set Approach

A crucial part in surface slab calculations is the convergence of the number of slab layers. There need to be enough layers below the surface to resemble bulk-like behavior, but as little as possible to reduce computational cost. In our case, more than 6 layers of silver are necessary for a sufficiently converged adsorption energy, which increased the computational cost to an intractable level. Fortunately, a fully accurate quantum mechanical description is only necessary for the interaction area of a molecule and the substrate. Thus, the fact that FHI-aims[1–5] uses atom-centered basis sets was utilized in the following way: The three uppermost layers of the surface slab were calculated with a tight and thoroughly converged basis, while the lower five layers (which are further away from the molecule) were only represented via a very light basis (see Supplementary Figure 1).

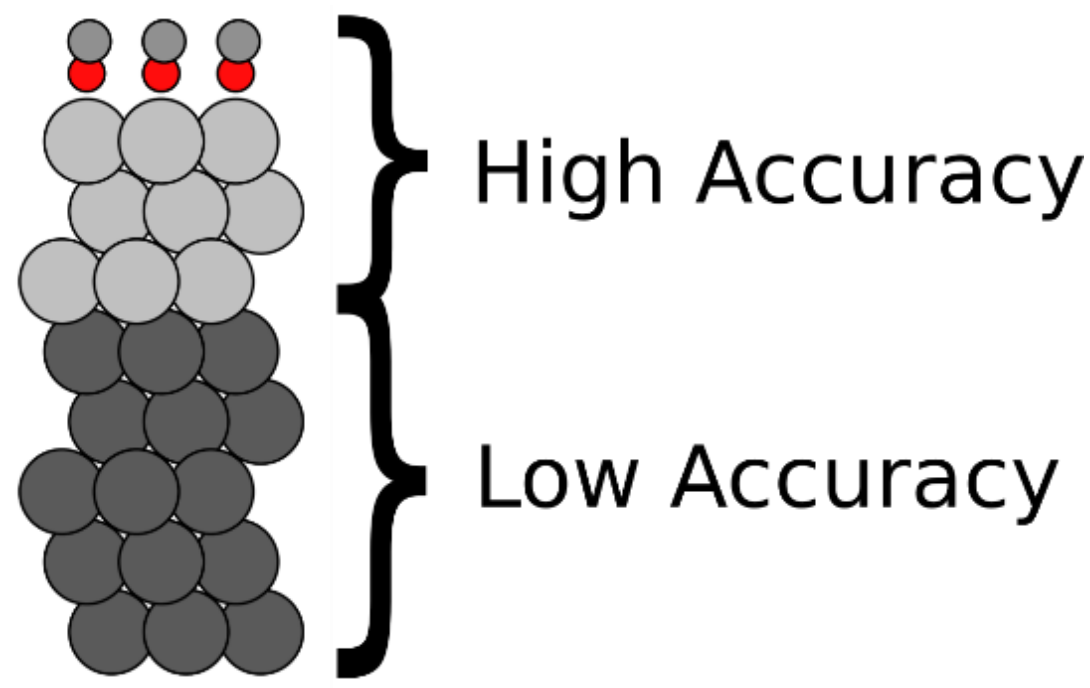

**Supplementary Figure 1: Visualization of CO adsorbed on the mixed basis used in this manuscript. The light grey layers consist of Ag atoms with tight basis while the dark grey atoms were calculated with a much looser basis.**

## 1.2 Finding Adsorption Geometries

Gaussian process regression[6] was used to interpolate the adsorption geometries of single molecules on a multidimensional potential energy surface, in the following way: First, an individual molecule was optimized lying flat on the surface (as was also found experimentally for P2O)[7] at an unspecific in-plane position to estimate the general distortions of the molecule upon adsorption on a surface. This optimized molecule was symmetrized and lifted from the surface by 0.1 Å to reduce the influence of the Pauli repulsion, i.e. to avoid “ramming” the molecule into the substrate. The lateral position of minima is not affected by the slightly larger vertical distance. This approach resulted in a mean distance between the center of the uppermost Ag layer and the molecular backbone of 2.4 Å for B2O, 2.9 Å for A2O and 3.0 Å for P2O. Then, a full potential energy surface (PES) for movement of this molecule along the x-y plane and rotation around the z-axis was mapped by performing approx. 50 DFT calculations. Subsequently, a geometry optimization for the full molecule and substrate was performed from all extrema in the aforementioned PES, allowing the whole molecule to relax until the remaining forces were below a suitable threshold for all atoms (0.02 eV/Å for A2O and P2O, 0.05 eV/Å for B2O). Those adsorption geometries were then symmetrized according to the applicable substrate symmetries to remove geometry-optimization artefacts. All final adsorption geometries, where at least one atom position differed by more than 0.1 Å (with symmetries taken into account), were considered as separate adsorption geometries.

## 1.3 Generating Motif Candidates

For all three molecules in this study we varied the number of molecules per cell $N_A$ from 1 to 4 and varied the unit cell size from $N_A$ * $A_{\min}$ to $N_A$ * ($A_{min}$ + 5) to ensure that experimentally feasible configurations are part of the prediction set. Here $A_{min}$ is the minimal cell size where configurations could be built without interfering with minimal distance thresholds set in SAMPLE.

For B2O and A2O we additionally constructed tightly packed polymorphs with up to 6 molecules per unit cell. The hexagonal cell for A2O was found by constructing all configurations for hexagonal unit cells with six molecules per unit cell up to 90 surface atoms per cell.

## 1.4 Hyper Parameters Used for SAMPLE

For the SAMPLE approach[8] several hyper parameters are necessary. All of those hyper parameters were varied systematically to maximize the log-likelihood in the Bayesian linear regression formalism. Supplementary Table 1 contains all optimized hyper parameters used for the prediction of the three systems. Decay length differences arise from the different molecule sizes. To avoid the need to fit the highly repulsive Pauli repulsion region with Bayesian linear regression, SAMPLE uses a minimal distance threshold for all atom-species combinations. For this work the following thresholds were used: O↔H: 1.6 Å; O↔O: 2.4 Å, H↔H: 1.6 Å; C↔H: 2.3 Å; C↔O: 2.5 Å.

**Table S1**: Hyper parameters of the SAMPLE approach used for the structure prediction of benzo-, anthra- and pentacenequinone.

| Hyper parameter | B2O | A2O | P2O |
|---|---|---|---|
| Adsorption energy uncertainty | 100 meV | 100 meV | 100 meV |
| Interaction energy uncertainty | 300 meV | 300 meV | 300 meV |
| DFT data uncertainty | 10 meV | 5 meV | 5 meV |
| Decay length | 5 Å | 5 Å | 10 Å |
| Decay power | 3 | 3 | 3 |
| Decay length feature space | 12 | 9 | 5 |
| Feature threshold | 0.0075 | 0.0075 | 0.01 |

# 2 Supporting Methodology (Experiment)

## 2.1 Details of the FFT Fit Procedure

Since the LEED device used by us typically probes a surface area in the order of 1 $mm^2$, the resulting images usually show a superposition of all motifs present in that area, e.g., symmetrically equivalent domains and other polymorphs, if present. For this reason, we additionally performed a detailed analysis of the STM images to ensure the assignment of the epitaxial relations determined by LEED to the motifs presented in STM as well as to increase the accuracy. We found motifs exhibiting Moiré patterns in their respective STM images on all samples presented in this study. These patterns show up as discrete spots in the two-dimensional Fourier transforms (FFT) as well, and their spot positions can be described in the same way as multiple scattering in geometric scattering theory, thus enabling an analysis[9] with LEEDLab[10]. For that purpose, we subjected the STM images showing a single domain of the motif under investigation, featuring molecular resolution as well as a Moiré pattern, to an FFT. We then determined the epitaxial relation by fitting the respective reciprocal lattice including the Moiré spots to the FFT, optimizing the adsorbate lattice and substrate lattice simultaneously, thus circumventing the distortions typically present in STM. This procedure yielded the epitaxy matrices for B2O and P2O directly (see Figure S2). However, it is not applicable to the hexagonal A2O structure, where we observed no Moiré patterns (likely due to its commensurate registry). Instead, we analyzed an STM image containing the motif presented in the main text (Motif A) and an additional, non-commensurate motif (Motif B), which shows a Moiré contrast, as can be seen in Figure S3. By taking a detail of the STM image containing only Motif B, we determined its lattice vectors relative to the substrate vectors, which then contain the local distortion, using the procedure described earlier. We presume the distortions to be constant for the image as a whole and therefore particularly for a detail containing only Motif A and use this detail to determine the lattice vectors of Motif A relative to Motif B, this time solely optimizing the adsorbate lattice. We then calculated the epitaxial relation of Motif A relative to the substrate by combining the two derived epitaxy matrices via matrix multiplication. By comparing both obtained motifs in reciprocal space to the LEED image of the A2O sample (Figure S4), it is evident that a structure determination based on LEED alone would remain ambiguous due to the similarity in many spot positions. Motif B is likely a kinetically trapped structure and could thus, and due to its non-commensurate nature, not be found within the theoretical framework (For details see Supporting Section 3).

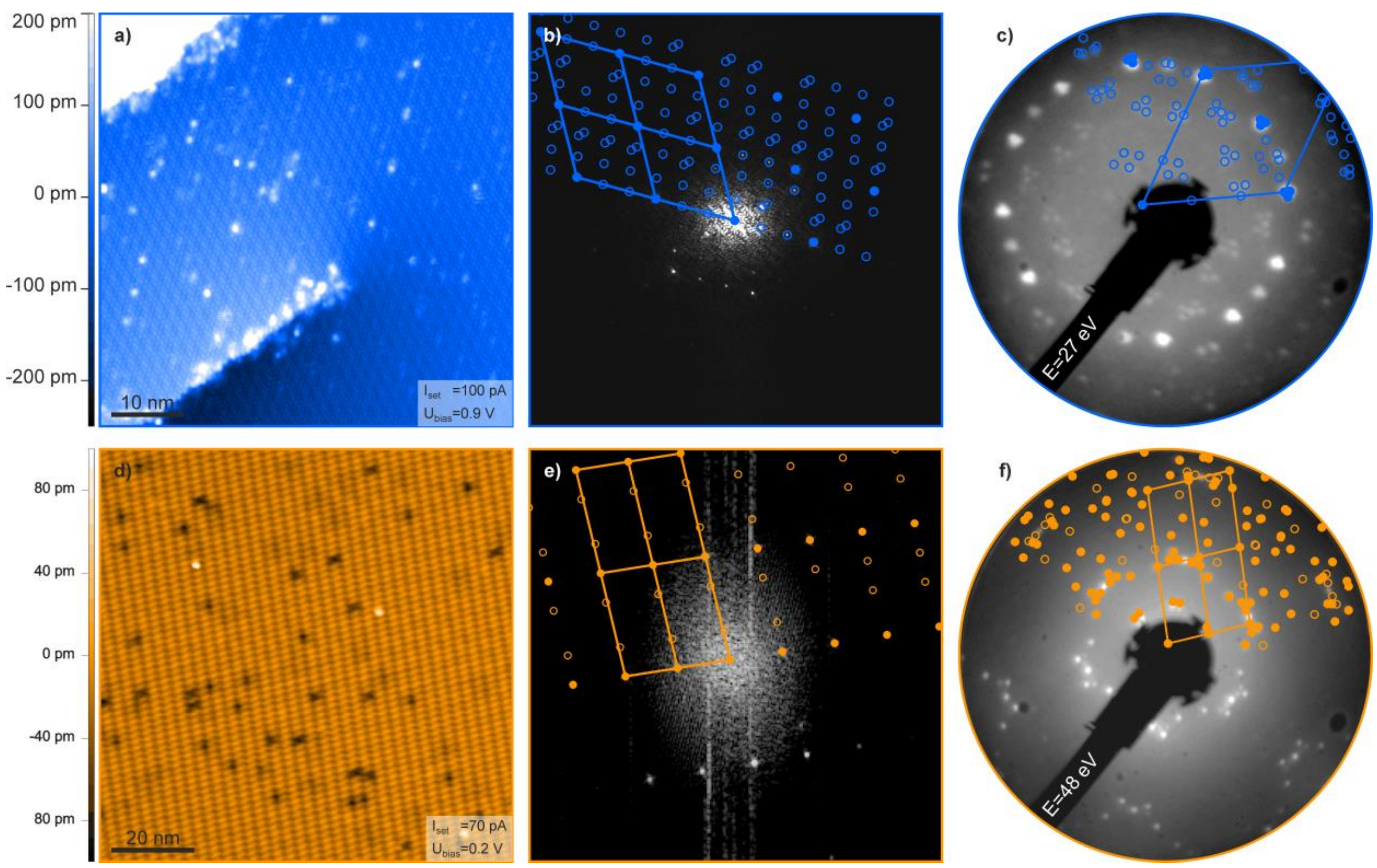

**Figure S2: STM images of a) B2O (blue) and d) P2O (orange) (same sample as the images in Figure 4) b,e) FFTs of a) and d), respectively, superimposed with the fitted reciprocal lattice. Several lattice points (dots) as well as Moiré frequencies (circles) are highlighted. c,f) LEED images (same measurement as Figure 4) at a given primary electron energy E superimposed with a simulation of the reciprocal lattice (dots) including multiple scattering (circles) and symmetrically equivalent domains as fitted to (b) and (e), respectively.**

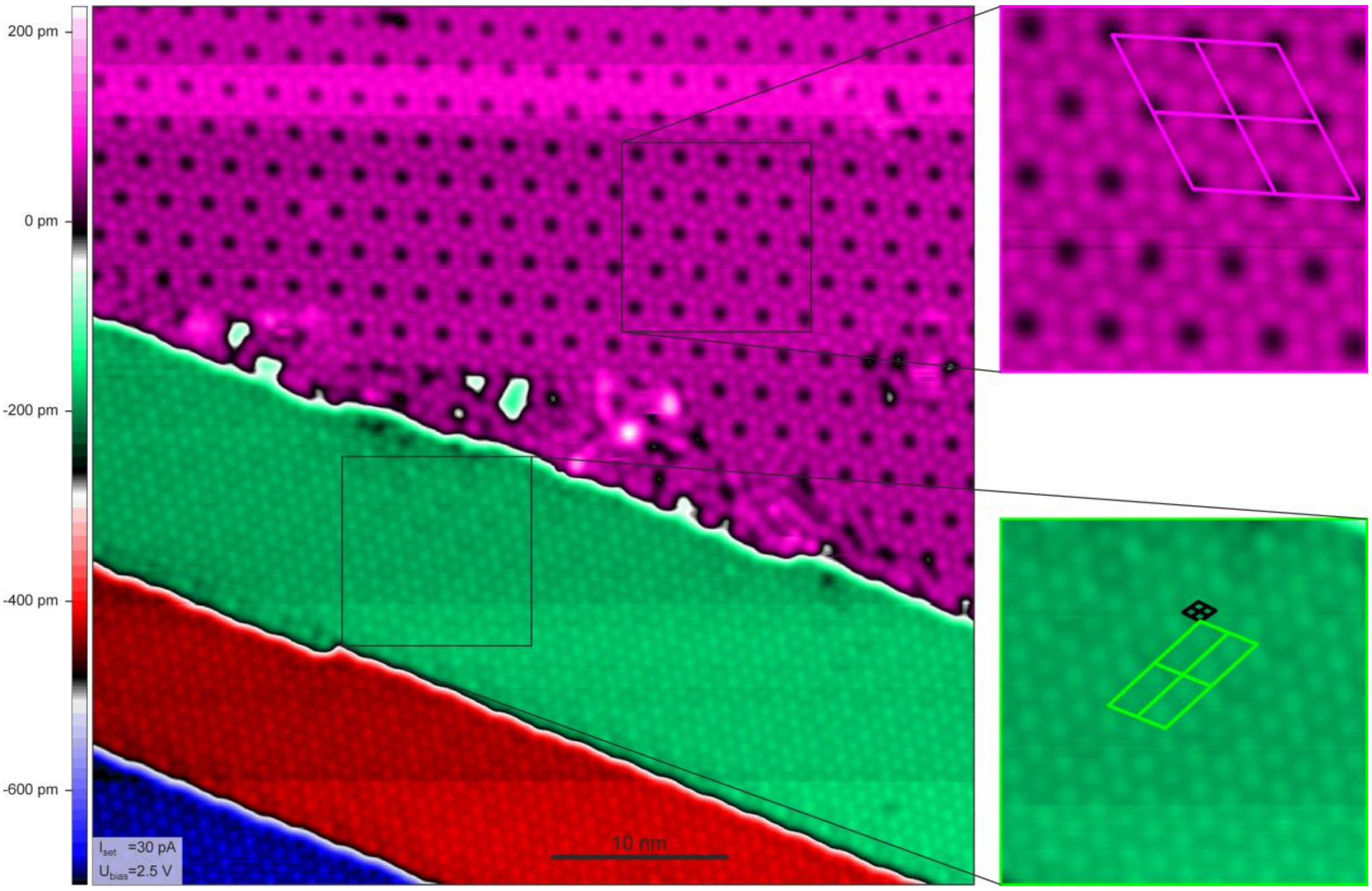

**Figure S3: STM image used for the structure analysis of A2O. Details of Motif A (top, magenta color code) and Motif B (bottom, green color code) feature representations of the unit cells of the motifs fitted to an FFT of the respective detail. The substrate lattice characterizing the distortions present in the image as determined by the fit is shown in comparison to Motif B (black lattice cell).**

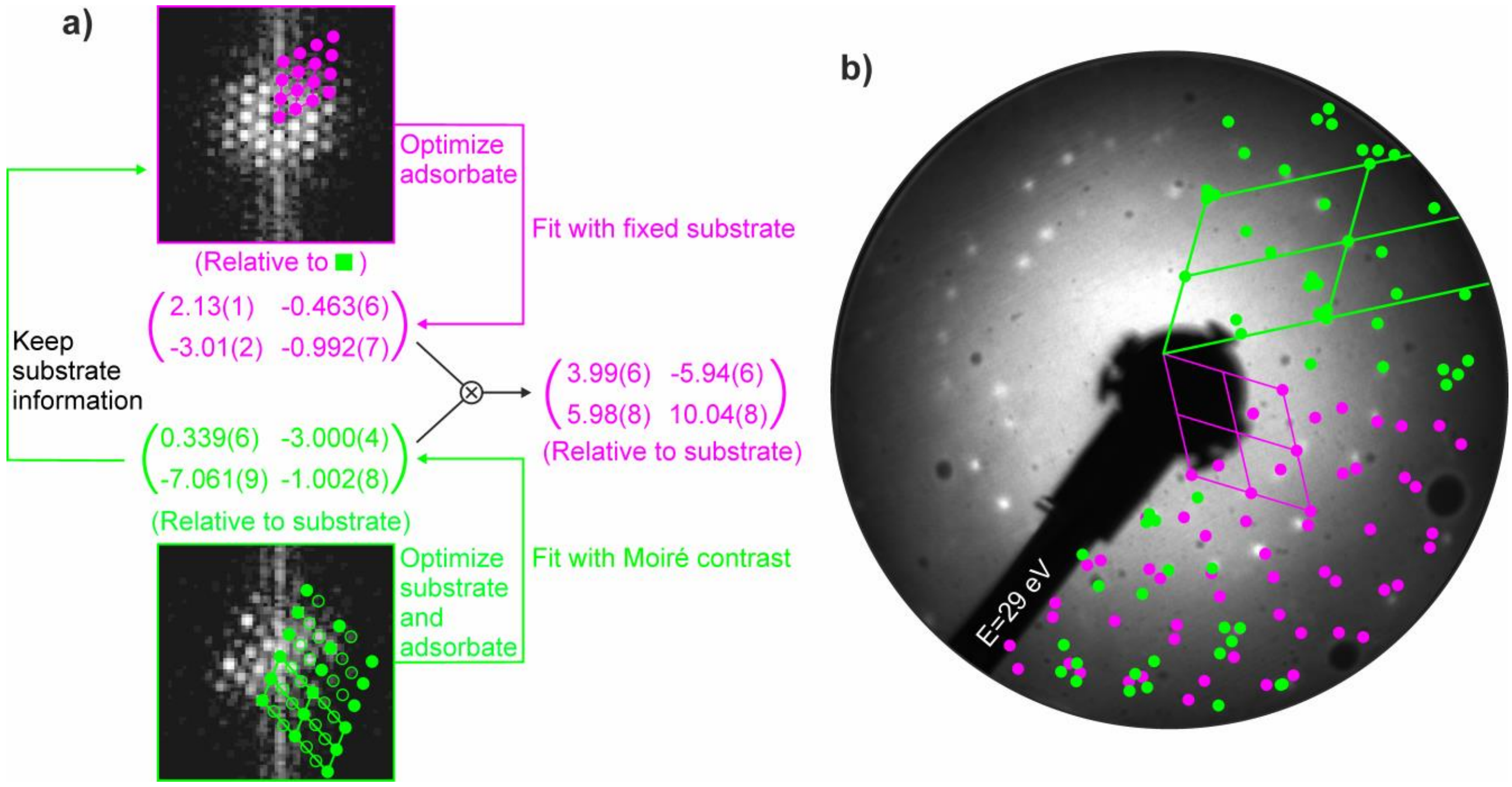

**Figure S4: Reciprocal space representation of the lattices of Motif A (magenta) and Motif B (green).a) FFTs of the details displayed in Figure S3 superimposed with the reciprocal unit cell as fitted to the FFT including the respective epitaxy matrix. Several reciprocal lattice points (filled circles) as well as Moiré spots (open circles) for Motif A and B are displayed as guide to the eye. b) LEED of the sample measured at 296 K, superimposed with a simulation of the reciprocal lattice points including symmetrically equivalent lattice representations.**

# 3 Predictions of All Possible Motifs

To obtain the motifs discussed in the main paper, first all possible motifs within the model discretization were created (see Supporting Section 1.4) to then predict all their energies. The energy models used for prediction were trained on a representative number of D-optimally chosen calculations (249 for B2O, 245 for A2O and 84 for P2O). To speed up training, intermolecular interaction data obtained from free-standing monolayers (i.e., removed substrate) was used as prior for the on-surface models. The prediction results and representative, energetically favorable motifs are presented in Figures S5 to S7.

The reliability of the model was verified via leave-one-out cross validation (LOOCV) uncertainties of 11 meV for B2O, 9 meV for A2O and 20 meV for P2O. The LOOCV uncertainty is obtained by calculating the root mean square error for all data points, but instead of using a separate test set, the deviation for each data point is obtained by training on all points but the one for which the deviation is probed. This, to some extent, is a worst-case measure of the uncertainty, as, if there are important data points in the set, the uncertainty for those will be very large, increasing the overall uncertainty. Due to the fact that our points are chosen D-optimally, meaning that the most important points should be calculated, we consider the LOOCV uncertainty superior to simple RMSE evaluation on a separate test set, additionally saving the costs of calculating an expensive test set.

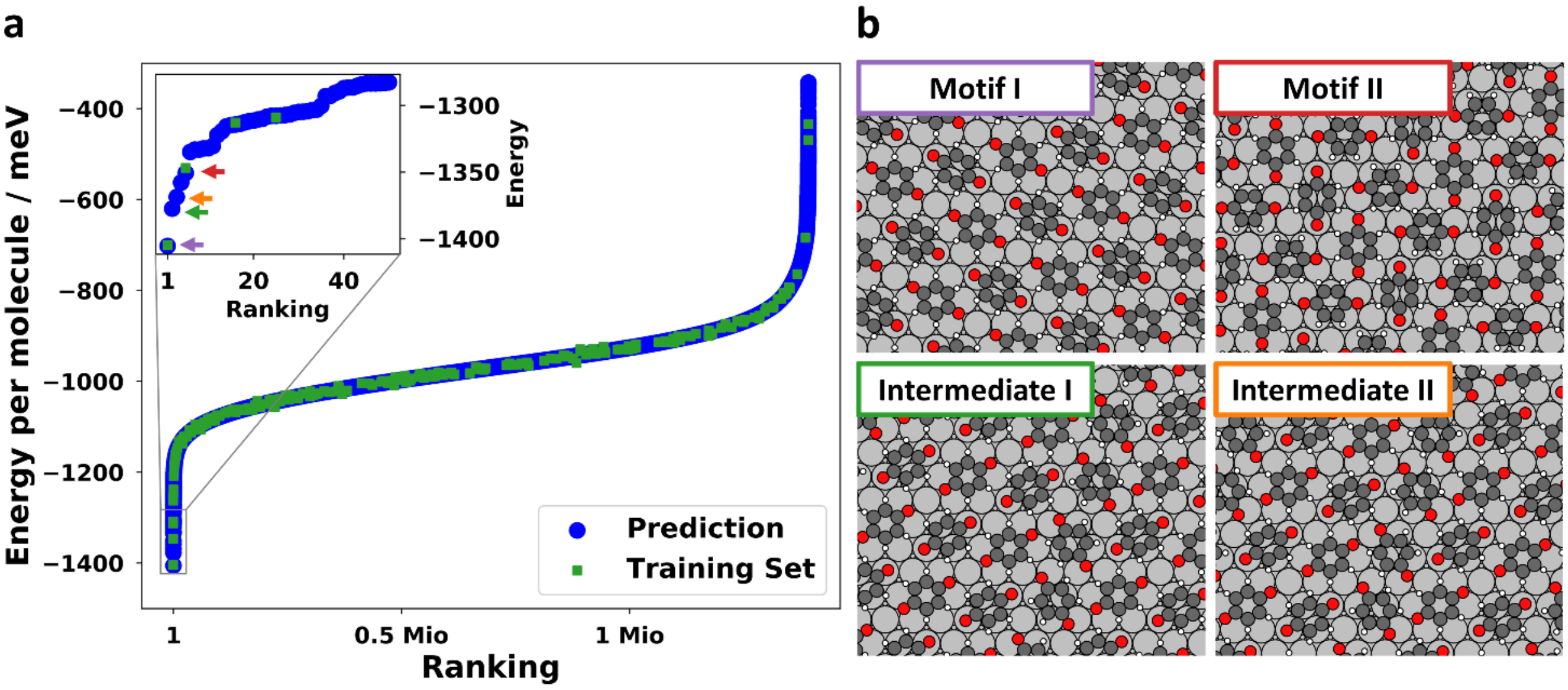

**Figure S5: Prediction and best theoretical motifs for B2O. a) Visualization of the formation energies per molecule for all predicted polymorphs of B2O and ranking according to energy from most favorable to least favorable. Points used for the training of the respective models are visualized with green rectangles. b) Motifs which were used for comparison to experiment and two intermediate motifs that rank between them.**

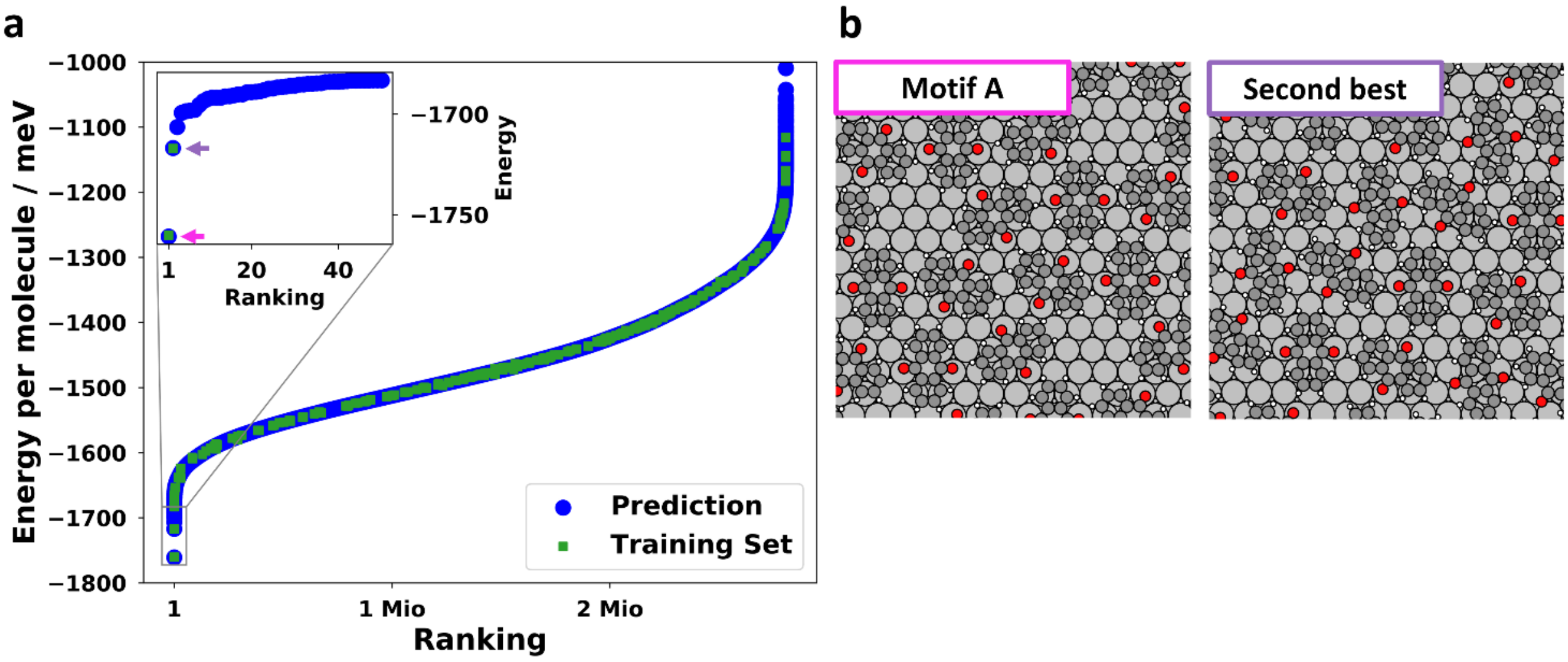

**Figure S6: Prediction and best theoretical motifs for A2O. a) Visualization of the formation energies per molecule for all predicted polymorphs of A2O and ranking according to energy from most favorable to least favorable. Points used for the training of the respective models are visualized with green rectangles. b) Best motif (corresponding Motif A in Figure S3) and second-best motif. Predictions of Motif B of Supporting Section 2.3 are beyond our theoretical framework due to its point-on-line epitaxy.**

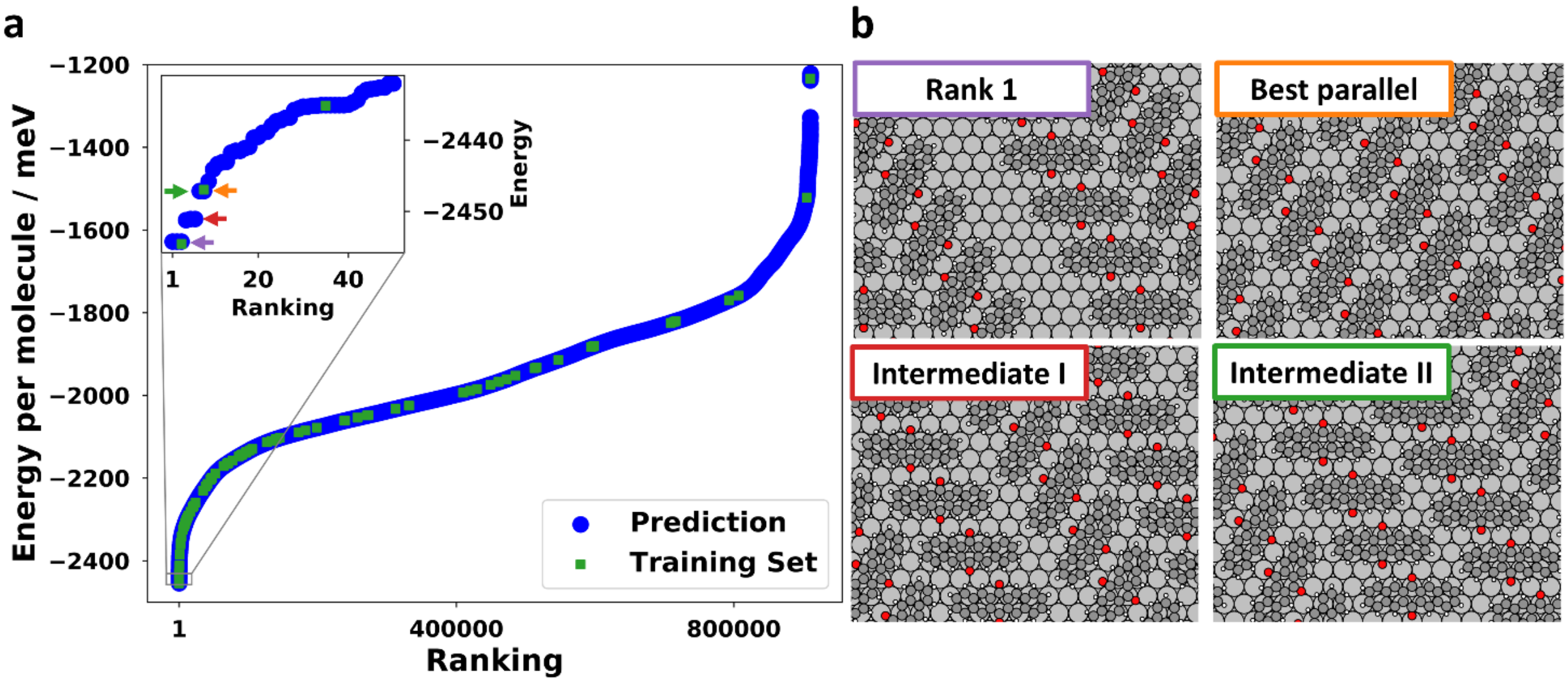

**Figure S7: Prediction and best theoretical motifs for P2O. a) Visualization of the formation energies per molecule for all predicted polymorphs of P2O and ranking according to energy from most favorable to least favorable. Points used for the training of the respective models are visualized with green rectangles. b) Visualization of the motifs that are predicted energetically slightly more beneficial (< 20 meV) than the experimental motif (Best parallel).**

# 4 Unit Cell Comparison

Table 1 shows the most important unit cells presented in Figure 4, a thorough analysis of all presented unit cells and uncertainties of the epitaxy matrices is given in Table S2. A graphical representation of the unit cell parameters used for Table 1 and Table S2 is presented in Figure S9. For the calculation of the lattice vector lengths in Table S2 we used the converged lattice constant of our calculations, which amounts to 4.019 Å for the conventional unit cell (corresponding to a minimal Ag-Ag distance of 2.842 Å).

The experimental uncertainties of the matrix elements indicate the simple standard deviations of the values, while each value represents the result of a directly optimized parameter of the fit routine used. Only the matrix elements for A2O are derived by the combination of two different matrices (as shown in Figure S4). Therefor the uncertainties of this matrix are obtained via Gaussian error propagation as well as the uncertainties of the lattice parameters.

**Table S2**: Comparison of experimental (grey background) and theoretical unit cells.
$a_1$, $a_2$: Lengths of lattice vectors derived from the epitaxy matrices in combination with the theoretical substrate lattice vectors; Γ: angle of the unit cell; Θ: angle between $a_1$ and the primitive substrate axis. Experimental uncertainties are indicated with brackets.

| | $a_1$ [Å] | $a_2$ [Å] | Γ [°] | Θ [°] | A [Å²] | Epitaxy |
|---|---|---|---|---|---|---|
| B2O I small | 6.875(5) | 6.603(3) | 119.88(5) | -14.28(2) | 39.36(4) | $\begin{pmatrix} 2.000(1) & -0.689(1) \\ 0.667(1) & 2.584(1) \end{pmatrix}$ |
| | 6.512 | 6.512 | 98.21 | 10.89 | 41.97 | $\begin{pmatrix} 2.5 & 0.5 \\ 0.5 & 2.5 \end{pmatrix}$ |
| B2O I large (Motif I) | 11.926(8) | 13.21(1) | 91.21(8) | 14.40(5) | 157.5(2) | $\begin{pmatrix} 4.667(2) & 1.205(3) \\ 1.334(2) & 5.169(3) \end{pmatrix}$ |
| | 13.024 | 13.024 | 98.21 | 10.89 | 167.87 | $\begin{pmatrix} 5 & 1 \\ 1 & 5 \end{pmatrix}$ |
| B2O II small | 6.875(5) | 6.603(3) | 119.88(5) | -14.28(2) | 39.36(4) | $\begin{pmatrix} 2.000(1) & -0.689(1) \\ 0.667(1) & 2.584(1) \end{pmatrix}$ |
| | 6.512 | 7.211 | 121.07 | -10.89 | 40.22 | $\begin{pmatrix} 2 & -0.5 \\ 0.5 & 2.75 \end{pmatrix}$ |
| B2O II large (Motif II) | 13.75(1) | 11.448(6) | 88.51(7) | -14.30(3) | 157.4(2) | $\begin{pmatrix} 4.000(3) & -1.380(2) \\ 3.334(2) & 4.476(2) \end{pmatrix}$ |
| | 13.024 | 12.388 | 94.31 | -10.89 | 160.88 | $\begin{pmatrix} 4 & -1 \\ 3 & 5 \end{pmatrix}$ |
| A2O | 24.6(3) | 24.9(2) | 120(1) | -36.5(4) | 530(10) | $\begin{pmatrix} 3.99(6) & -5.94(6) \\ 5.98(8) & 10.04(8) \end{pmatrix}$ |
| | 24.776 | 24.776 | 120.00 | -36.57 | 531.60 | $\begin{pmatrix} 4 & -6 \\ 6 & 10 \end{pmatrix}$ |
| P2O | 14.96(2) | 8.156(4) | 95.92(8) | 40.78(3) | 121.3(2) | $\begin{pmatrix} 2.000(1) & -3.969(6) \\ 3.000(1) & 2.719(2) \end{pmatrix}$ |
| | 15.038 | 8.526 | 100.89 | 40.89 | 125.90 | $\begin{pmatrix} 2 & -4 \\ 3 & 3 \end{pmatrix}$ |

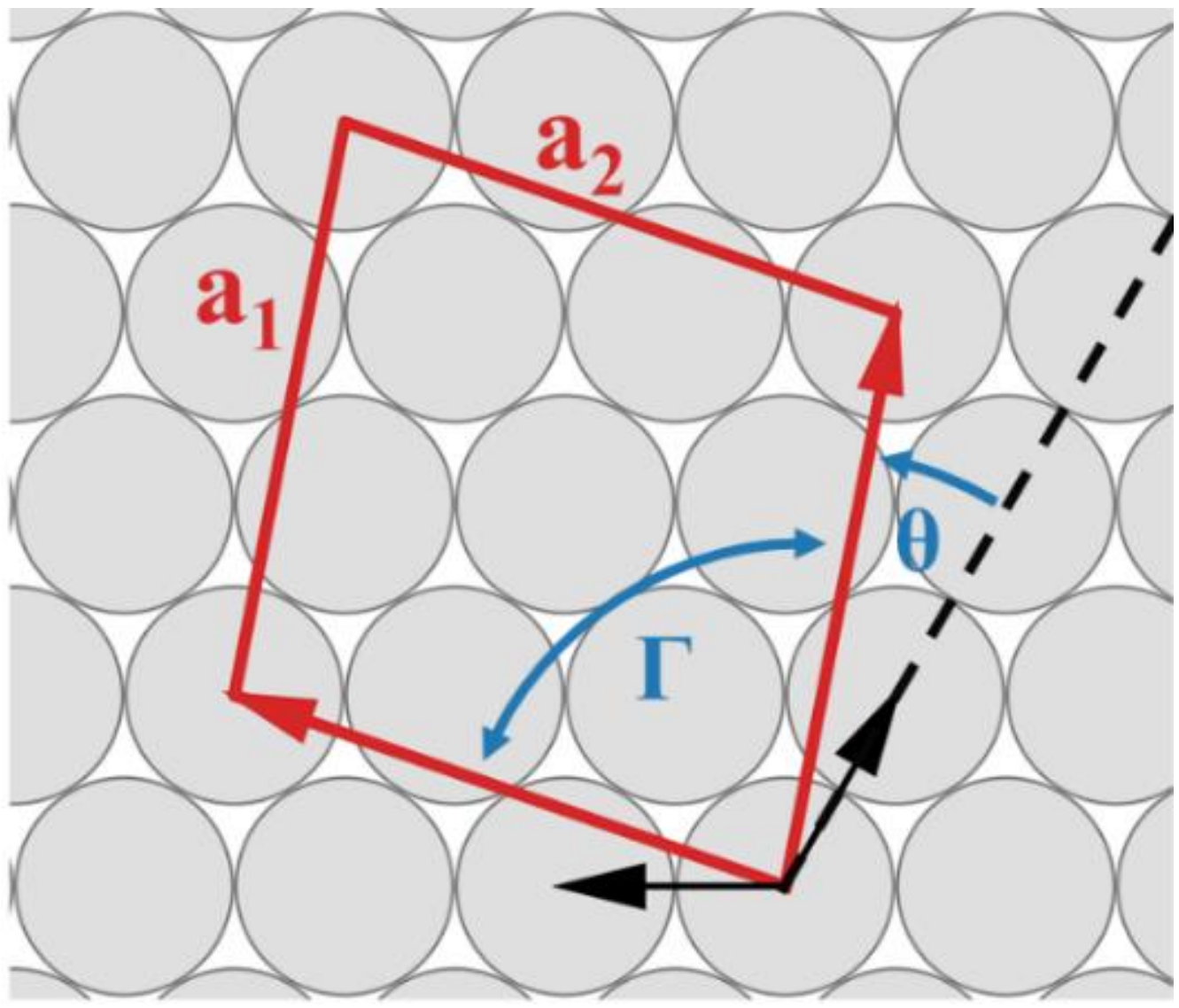

**Figure S9: Visual explanation of the unit cell parameters used for cell comparison. $a_1$ and $a_2$ are the lengths of the unit cell lattice; Γ is defined as the enclosing angle of the unit cell vectors; Θ represents the angle between the first lattice vector and a primitive substrate axis, negative values correspond to angles in anticlockwise direction.**

# 5 Additional Pair Potential Plots

Figure 3 visualized interactions of pairs of molecules with parallel alignment. Figure S10 presents a selection of interaction maps for pairs with non-equal orientation.

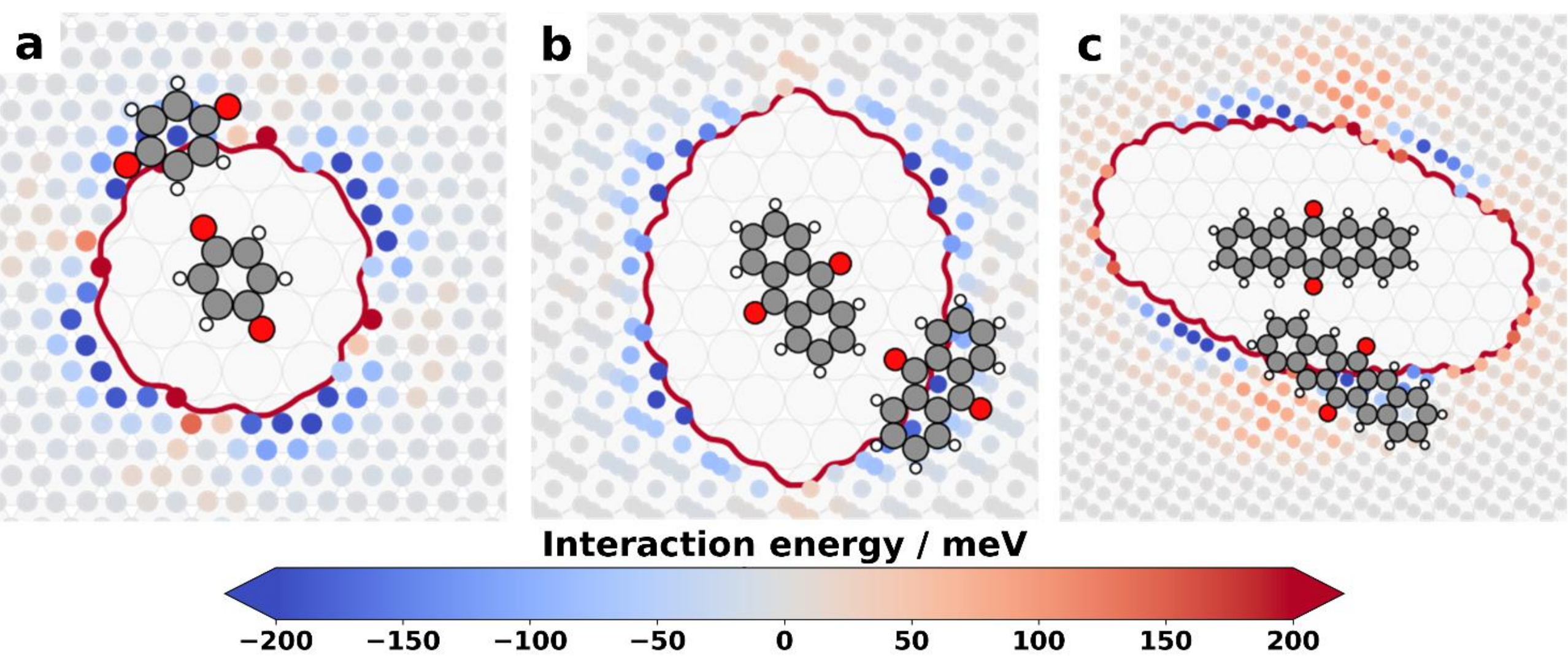

**Figure S10: Total pairwise interaction energies for non-equivalent molecular orientation. Each circle represents a possible pairwise interaction between the central molecule and an adjacent molecule centered at the circle position for (a) B2O, (b) A2O and (c) P2O. The red contour shows the minimal distance before a pair is considered colliding. The circle color indicates the corresponding interaction energy.**